\documentclass[journal]{IEEEtran}

\IEEEoverridecommandlockouts

\usepackage{enumitem}
\usepackage{graphicx}
\usepackage{algorithmic}
\usepackage{epsfig}
\usepackage{wrapfig}
\usepackage{amsmath,amsthm, amssymb}
\usepackage[ruled,linesnumbered]{algorithm2e}
\usepackage{multirow}
\usepackage{multicol}
\usepackage{tabularx}
\usepackage[dvipsnames]{xcolor}
\usepackage{comment}
\usepackage{subcaption}

\usepackage{array}
\usepackage{color}
\usepackage{hyperref}
\usepackage{breqn}
\usepackage{cite}
\newcommand{\N}{\mathcal{N}}
\newcommand{\E}{\mathcal{E}}
\newcommand{\G}{\mathcal{G}}

\newcommand{\A}{\mathcal{A}}

\makeatletter
\let\old@ps@headings\ps@headings
\let\old@ps@IEEEtitlepagestyle\ps@IEEEtitlepagestyle
\def\psccfooter#1{%
    \def\ps@headings{%
        \old@ps@headings%
        \def\@oddfoot{\strut\hfill#1\hfill\strut}%
        \def\@evenfoot{\strut\hfill#1\hfill\strut}%
    }%
    \def\ps@IEEEtitlepagestyle{%
        \old@ps@IEEEtitlepagestyle%
        \def\@oddfoot{\strut\hfill#1\hfill\strut}%
        \def\@evenfoot{\strut\hfill#1\hfill\strut}%
    }%
    \ps@headings%
}
\makeatother

\begin{document}
%
% paper title
% Titles are generally capitalized except for words such as a, an, and, as,
% at, but, by, for, in, nor, of, on, or, the, to and up, which are usually
% not capitalized unless they are the first or last word of the title.
% Linebreaks \\ can be used within to get better formatting as desired.
% Do not put math or special symbols in the title.
\title{Distributed Computing for Scalable Optimal Power Flow in Large Radial Electric Power Distribution Systems with Distributed Energy Resources}

%% To specify the authors when (number of affiliations <= 2)
\author{
\IEEEauthorblockN{Rabayet Sadnan$^{*}$ and Anamika Dubey}\\
% \vspace{-10pt}
\IEEEauthorblockA{{Department of Electrical Engineering and Computer Science,}
{Washington State University} \\
Email: \{rabayet.sadnan$^*$, anamika.dubey\}@wsu.edu}
\vspace{-11pt}}

% make the title area
\maketitle

% As a general rule, do not put math, special symbols or citations
% in the abstract
\begin{abstract}
Solving the non-convex optimal power flow (OPF) problem for large-scale power distribution systems is computationally expensive. An alternative is to solve the relaxed convex problem or linear approximated problem, but these methods lead to sub-optimal or power flow infeasible solutions. In this paper, we propose a fast method to solve the OPF problem using distributed computing algorithms combined with a decomposition technique. The full network-level OPF problem is decomposed into multiple smaller sub-problems defined for each decomposed area or node that can be easily solved using off-the-shelf nonlinear programming (NLP) solvers. Distributed computing approach is proposed via which sub-problems achieve consensus and converge to network-level optimal solutions. The novelty lies in leveraging the nature of power flow equations in radial network topologies to design effective decomposition techniques that reduce the number of iterations required to achieve consensus by an order of magnitude.
\end{abstract}


% Use this to place sponsorships
% \thanksto{\noindent Submitted to the 22nd Power Systems Computation Conference (PSCC 2022).}

\section{Introduction}
The nature and the requirements of the power systems, specially at the distribution-level is changing rapidly with a large-scale integration of controllable distributed energy resources (DERs). The continued proliferation of DERs, that include Photovoltaic (PV) systems, battery energy storage units (BESS), and controllable loads such as Electric Vehicles (EVs) is leading to a drastic increase in the number of active nodes at the distribution-level that need to be controlled/managed optimally for an efficient and resilient grid operations. Traditionally, grid operations are centrally managed upon solving an optimal power flow (OPF) problem where centralized optimization techniques are used to solve the resulting difficult non-linear non-convex OPF problem \cite{momoh1999review1,castillo2013survey}. Unfortunately, the computational challenges, primarily posed by the non-convex power flow constraints in OPF formulation, increases drastically with the size of the distribution systems motivating computationally efficient approaches \cite{Jha2019}. 

Existing methods manage the computational challenges using convex relaxation or linear approximation techniques \cite{gan2014convex, Jha2021}. The primary drawbacks of the convex relaxed models are the possibilities of inexact/infeasible power flow solutions \cite{JhaPV}. The linear approximated models may lead to NLP infeasible solutions and high optimality gap depending upon the problem type. Moreover, methods based on both approximation and relaxation techniques use a centralized paradigm that may lead to scalability challenges as the problem size increases. With a majority of DER integration happening at the secondary feeder level, the OPF problem will need to solve even larger feeder with thousands of secondaries. For example, the largest IEEE test feeder is a 8500 node test system that terminates at the secondary transformer level and does not include secondary feeders. If each service transformer is expanded to a 20 node secondary feeder, it will lead to a total of 22000 secondary nodes added to the problem formulation. Such problem complexities motivate the move towards a distributed computing paradigm. In general, the distributed computing is facilitated by the rapid growth in high-performance computing using many-core machines. Fortunately, the radial operational topology of power distribution systems make it highly conducive for parallelization and distributed computing, In this paper, we develop a distributed computing approach for OPF problems for distribution systems, that can scale for very large distribution feeders and converges using fewer message passing among distributed computing nodes thus significantly reducing the overall compute time.

Within this context, existing literature includes numerous approaches on the application of distributed optimization algorithms for power distribution systems \cite{molzahn2017survey, peng2016distributed}. In general, these methods adopt the traditional distributed optimization techniques to model a distributed optimal power flow (D-OPF) problem  \cite{molzahn2017survey,boyd2011distributed, zheng2015fully,peng2016distributed}. A D-OPF formulation decomposes the OPF into several smaller subproblem that require multiple micro- and macro-iterations for convergence. Micro-iterations involve solving the distributed sub-problems in parallel. And macro-iterations involve exchanging the solutions or more specifically the updated boundary variables obtained from the distributed subproblems. Both micro and macro-iterations together decide the time-of-convergence (ToC) for the algorithm. Unfortunately, the exiting distributed optimization algorithms require a very large number of macro-iterations to converge for medium-scale distribution grids \cite{erseghe2014distributed, dall2013distributed, peng2016distributed, lu2017fully}. A practical implementation of such algorithms requires a very fast communication among distributed computing nodes to reach a converged solution within a reasonable time. A large number of communication rounds/message-passing events among distributed agents is not preferred since this leads to significant delays in decision-making. Lately, to address some of these challenges, real-time feedback based online distributed algorithms have been explored in the related literature for network optimization \cite{ cavraro2017local,bernstein2019real,bastianello2020distributed,qu2019optimal,hu2019branch,magnusson2020distributed}. Generally, these algorithms do not wait to optimize for a time-step but asymptotically arrive at an optimal decision over several steps of real-time decision-making. However, these algorithms also take hundreds of iterations to converge/track the optimal solution for a mid-size feeder. This raises further challenges to the performance of the algorithm for larger feeders, especially during the fast-varying phenomenon and slow communication channels. 

To address these challenges, recently we have developed a distributed OPF formulation for the radial distribution systems based on the equivalence of networks principle \cite{sadnan2020real,sadnan2021distributed}. In this paper, we further expand the previously proposed method to solve OPF for very large notional distribution test feeders (with 10,000 nodes) for several different problem objectives. The proposed approach solves the original non-convex optimal power flow problem for power distribution systems using a novel decomposition technique combined with distributed computing approach. The distributed subproblems are related via the boundary variables shared by the neighboring nodes. First, the low-compute distributed OPF sub-problems are locally solved. The consensus of the boundary variables is achieved using a Fixed-Point Iteration (FPI) algorithm. Upon consensus, the solutions converge to network-level OPF solutions. The proposed approach leverages the radial topology of the power distribution system and the associated unique power flow properties to develop message passing routines that reduces the number of message-passing among distributed agents by an order of magnitude. We demonstrate the proposed approach for three problem objectives (1) loss minimization (2) DER generation maximization and (3) voltage deviation minimization using a single-phase equivalent of 8500-node test feeder (with 2500 nodes) and a balanced synthetic 10,000 node distribution feeder. The proposed approach is shown to scale for all problem objectives while most centralized formulation can’t be solved for more than 2000 nodes using off-the-shelf optimization solvers such as \textit{Artelys Knitro}.  To our knowledge, this is the first paper to demonstrate an approach that solves such a large-scale D-OPF on a regular CPU without the use of any high-performance computing (HPC) machines. 

%%%%  Central OPF problem formulation Section %%%%%%%%%
\section{Centralized OPF Model}
In this paper, $(\cdot)^*$ represents the complex-conjugate; $(\cdot)^T$ represents matrix transpose; $|~.~|$ symbolizes the absolute value of a number or the cardinality for a discrete set; $(\cdot)^n$ represents the $n^{th}$ iteration;
and $j = \sqrt{-1}$ in a complex number; $\overline{(.)}$ and $\underline{(.)}$  denotes the maximum and minimum limit of a given quantity.

\subsection{Network and DER models}
Let us consider a balanced radial power distribution network -- represented by the directed graph $\G = (\N, \E)$, where $\N$ be the set of all nodes in the system and $\E$ denotes the set of all distribution lines connecting the ordered pair of buses $(i,j)$ i.e., from node $i$ to node $j$. Also, $r_{ij}+jx_{ij}$ is the series impedance $\forall\{ij\}\in \E$. Let, for node $j$, $k$ be the set of all children nodes; thus, in $k: j\rightarrow k$, $k$ represents the set of children nodes for the node $j$. Next, we denote $v_j = |V_j|^2 = V_j{V_j}^*$ as the squared magnitude of voltage at node $j$. Let $l_{ij}$ be the squared magnitude of current flow in branch $\{ij\}$. We denote $P_{ij}, Q_{ij}$ as the sending-end active and reactive power flows for branch ${ij}$, and complex power $S_{L_j} = p_{L_j}+jq_{L_j}$ is the load connected and $S_{D_j} = p_{Dj}+jq_{Dj}$ is the power output of DER connected at node $j$. The network is modeled using the branch flow equations \cite{baran1989optimal1} defined for each line $\{ij\}\in \E$ and $\forall j \in \N$ in \eqref{pfmodel}.

 \vspace{-0.3cm}
\begin{small}
\begin{IEEEeqnarray}{C C}
\small
\IEEEyesnumber\label{pfmodel} \IEEEyessubnumber*
P_{ij}-r_{ij}l_{ij}-p_{L_j}+p_{Dj}= \sum_{k:j \rightarrow k} P_{jk}   \label{pfmodel1}\\
Q_{ij}-x_{ij}l_{ij}-q_{L_j}+q_{Dj}= \sum_{k:j \rightarrow k} Q_{jk} \label{pfmodel2}\\
v_j=v_i-2(r_{ij}P_{ij}+x_{ij}Q_{ij})+(r_{ij}^2+x_{ij}^2)l_{ij}\label{pfmodel3}\\
v_il_{ij} = P_{ij}^2+Q_{ij}^2 \label{pfmodel4}
\end{IEEEeqnarray}
\end{small}
\vspace{-0.3cm}

The DERs are modeled as Photovoltaic modules (PVs) interfaced using smart inverters, capable of two-quadrant operation. If the reactive power generation, $q_{Dj}$, is controllable and modeled as the decision variable for the optimal operation, then the real power generation by the DER, $p_{Dj}$, is assumed to be known (measured). Let the rating of the DER connected at node $j$ be $S_{DRj}$, then the limits on $q_{Dj}$ are given by \eqref{DG_lim}.

\vspace{-0.03cm}
 \begin{small}
\begin{equation}
\IEEEyesnumber\label{DG_lim}
-\sqrt{S_{DRj}^2-p_{Dj}^2} \leq q_{Dj} \leq \sqrt{S_{DRj}^2-p_{Dj}^2}
\end{equation}
 \end{small}
%  \vspace{-0.3cm}
On the contrary, if the active power generation, $p_{Dj}$, is modeled as the decision variable, then $q_{Dj}$ is set to $0$, and $p_{Dj}$ can vary between $0$ and $S_{DRj}$, see \eqref{DG_lim2}.

\vspace{-0.2cm}
\begin{small}
\begin{equation}
\IEEEyesnumber\label{DG_lim2}
0 \leq p_{Dj} \leq S_{DRj}
\end{equation}
 \end{small}
\vspace{-0.6cm}

\subsection{Centralized OPF problems}

To optimize the network for some cost function, we define a centralized OPF problem defined by a network-level problem objective, the power flow models in \eqref{pfmodel}, and the operating constraints on the power flow variables. In this paper, we formulate three different optimal power flow problems for the power distribution grids, (i) active power loss minimization, (ii) DER generation maximization, and (iii) Voltage deviation ($\Delta$V) minimization. The corresponding OPF problems are detailed below.

\subsubsection{Loss Minimization}
The problem objective is to reduce the network losses by controlling the reactive power output from DERs ($q_{Dj}$). Let $X_{lm}=[P_{ij}, ~Q_{ij}, ~l_{ij}, ~v_j, ~q_{Dj}]^T$ be the problem variables $\forall j\in \N$, and $\forall \{ij\}\in \E$. Note that, if node $j$ doesn't have any DER, then $q_{D_j} = 0$. Also, let $F_{lm}(X_{lm})$ denote the objective function representing the total power loss in the given distribution system. Note that $F_{lm}(X_{lm})$ is a function of both the power flow variables and decision variables. Then, the OPF problem is defined as the following in \textbf{(C1)}.

\vspace{-0.4cm} 
\begin{small}
\begin{IEEEeqnarray}{C C}
\IEEEyesnumber\label{LM_OPF} \IEEEyessubnumber*
\text{\textbf{(C1)}}\hspace{0.4cm} \min \hspace{0.2cm} F_{lm}(X_{lm}) = \sum_{\{ij\}\in \E} l_{ij}r_{ij} \hspace{0.6cm}\\
\text{s.t.}  \hspace{0.2cm} \text{\eqref{pfmodel} and \eqref{DG_lim}}\\
\underline{V}^2 \leq v_j \leq \overline{V}^2 \hspace{0.7cm} ;\forall j\in \N \label{LM_OPF1}\\
l_{ij} \leq \left(I^{rated}_{ij}\right)^2  \hspace{0.5cm} ;\forall \{ij\} \in \E \label{LM_OPF2}
\end{IEEEeqnarray}
\end{small}
\vspace{-0.4cm}

\noindent where, $\overline{V} = 1.05$ and $\underline{V} = 0.95$ are the limits on bus voltages, and $(I^{rated}_{ij})^2$ is the thermal limit for the branch $\{ij\}$.

\subsubsection{DER Maximization}
In DER maximization problem objective, the DER active power generation is maximized without violating the operational limits of the distribution system. This is achieved by maximizing the active power output from DERs ($p_{Dj}$). Let $X_{dm}=[P_{ij}, ~Q_{ij}, ~l_{ij}, ~v_j, ~p_{Dj}]^T$ be the problem variables. Here, the objective function is denoted by $F_{dm}(X_{dm})$, representing the total active power generation by DERs. Then, this DER maximization OPF problem is defined as the following in \textbf{(C2)}. Similar to the previous formulation, if any node $j$ doesn't have any DER, then we set $p_{D_j} = 0$.

\vspace{-0.4cm} 
\begin{small}
\begin{IEEEeqnarray}{C C}
\IEEEyesnumber\label{DM_OPF} \IEEEyessubnumber*
\text{\textbf{(C2)}}\hspace{0.4cm} \max \hspace{0.2cm} F_{dm}(X_{dm}) = \sum_{j\in \N} p_{Dj} \hspace{0.6cm}\\
\text{s.t.}  \hspace{0.2cm} \text{\eqref{pfmodel} and \eqref{DG_lim2}}\\
\underline{V}^2 \leq v_j \leq \overline{V}^2 \hspace{0.7cm} ;\forall j\in \N \label{DM_OPF1}\\
l_{ij} \leq \left(I^{rated}_{ij}\right)^2  \hspace{0.5cm} ;\forall \{ij\} \in \E \label{DM_OPF2}
\end{IEEEeqnarray}
\end{small}
\vspace{-0.4cm}

Kindly note that DER maximization problem is also known as PV hosting problem if DERs are modeled as PV modules.

\subsubsection{$\Delta$V Minimization}
In this specific network level optimization problem, we try to keep the nodal voltages as close as possible to a pre-specified reference, $V_{ref}$. The problem objective is to minimize the nodal voltage deviations from the reference value by controlling the reactive power output from DERs ($q_{Dj}$). The problem variables are denoted by $X_{dv}=[P_{ij}, ~Q_{ij}, ~l_{ij}, ~v_j, ~q_{Dj}]^T$, $\forall j\in \N$ and $\forall \{ij\}\in \E$. Also, the cost function, $F_{dv}(X_{dv})$, represents the total two-norm distances of nodal voltages, $v_j$, from reference voltage $v_{ref}$. Mathematically $F_{dv}(X_{dv}) = \sqrt{\sum(v_j-v_{ref})^2}$, $\forall j\in \N$. The OPF problem is defined as the following in \textbf{(C3)}. Here in this paper, we used $V_{ref} = 1.00$ as the bus reference voltage.

\vspace{-0.4cm} 
\begin{small}
\begin{IEEEeqnarray}{C C}
\IEEEyesnumber\label{DV_OPF} \IEEEyessubnumber*
\text{\textbf{(C3)}}\hspace{0.4cm} \min \hspace{0.2cm} F_{dv}(X_{dv}) = \sqrt{\sum_{\forall j\in \N}(v_j-v_{ref})^2} \hspace{0.6cm}\\
\text{s.t.}  \hspace{0.2cm} \text{\eqref{pfmodel} and \eqref{DG_lim}}\\
\underline{V}^2 \leq v_j \leq \overline{V}^2 \hspace{0.7cm} ;\forall j\in \N \label{DV_OPF1}\\
l_{ij} \leq \left(I^{rated}_{ij}\right)^2  \hspace{0.5cm} ;\forall \{ij\} \in \E \label{DV_OPF2}
\end{IEEEeqnarray}
\end{small}
\vspace{-0.4cm}

\noindent \textit{\textbf{Assumption 1:}} The loads in the network for all three OPFs are modeled as constant power loads; i.e., in \textit{ZIP} load model, $(Z, I, P) = (0, 0,1)$.

In the next section, we detail the method on how to decompose the optimization problems for large scale distribution grids into several sub-problems, solve in parallel, and converge into the final solution.

%%%%  Decomposition Section %%%%%%%%%
\section{Decomposition of the Central OPF Problem}
The OPF problems described in the previous section are formulated as a centralized optimization problem for the radial power distribution systems. For a large scale distribution system with thousands of nodes and decision variables, solving the NLP OPF is computationally expensive and difficult to converge for very large-scale distribution systems. Since the power distribution system is operated radially, the OPF problems defined in \textbf{(C1)-(C3)} are naturally decomposable into multiple sub-problems defined for the connected areas. The details of the proposed problem decomposition technique and the resulting distributed OPF problem are discussed next.

\begin{figure}[t]
     \vspace{-0.2cm}
    \centering
    \includegraphics[width=0.3\textwidth]{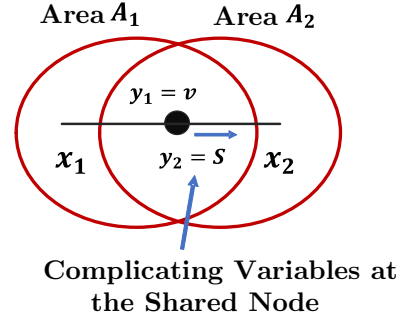}
%  \vspace{-0.3cm}
    \caption{An example of a two-area system.}
    \label{two_area_proposed}
 \vspace{-0.5cm}
\end{figure}

\subsection{Decomposition of the OPF Problem}
First, we decompose the whole distribution grid into $N$ smaller areas. Let $\A_R = \{A_1,~A_2, …,~A_{N}\}$, be the set of all decomposed areas. Also, let each area, $A_m \in \A_R $, be defined as a directed graph $A_m = \G(\N_m, \E_m)$. Here, each area $A_m$ has a maximum number of nodes/variables, so that the respective OPF sub-problems for that area can easily be solved using off-the-shelf NLP solvers. The coupling/complicating variables among these smaller sub-problems, associated with respective areas, are directed by the structure of the network. Since, the grid was radial to begin with, the decomposed areas, or the sub-trees of the networks are also connected radially with each other. This specific structure of the network helps to identify the unique parent area and the child areas for any area $A_m$, which in turns associates the complicating/shared variables -- exchanged among sub-problems to solve the overall master problem. For this decomposition method, the complicating variables are the shared bus voltages and power flows in the shared bus. Computationally, sub-problems associated with each area is solved in parallel by assuming a fixed voltage at the shared bus with the unique parent area, and a constant loads at the shared buses with child areas. After solving the sub-problems, the respective complicating variables, i.e., the total power requirements in that area is shared with sub-problem for the parent area and the shared bus voltages are shared with sub-problems associated with child areas. This exchange of complicating variables is called as macro-iteration here. After this macro-iteration, sub-problems are solved again, till convergence for the complicated variables are achieved.

%\noindent \textbf{\textit{Remark 1:}} {\color{red} Talk about why this assumption works for radial networks}.

Specifically, lets assume the network is decomposed into $2$ areas -- area $A_1$ and $A_2$; each with their own, purely local variables -- $x_1$ and $x_2$ (see Fig. \ref{two_area_proposed}). Area $A_1$ is the parent area of area $A_2$. Let $\textbf{Y} = [y_1, y_2]^T$ be the complicating variable that couples the sub-problems for the two areas. Here, $y_1$ and $y_2$ are the bus voltage magnitude ($v$) and the complex power flow through the bus ($S = P+jQ$) shared between $A_1$ and $A_2$, respectively; i.e., $[y_1, y_2]^T = [v, S]^T$. If the set of all local variables for $A_1$ and $A_2$ is denoted by $X_1$ and $X_2$, respectively, then $ X_1 = \{x_1, y_1\}{\hspace{0.2cm}} \text{\&{\hspace{0.2cm}}}X_2 = \{x_2, y_2\}$. Let $X = X_1 \cup X_2$ be the set of central OPF problem variables, and $\mathcal{S}$ is the set of constraints for the overall centralized optimization problem. If $F$ is a decomposable cost function, then the problem can be decomposed and written as \eqref{eqProposed_obj2}, where, $\mathcal{S}_1$ and $\mathcal{S}_2$ are the set of constraints on local variables for decomposed area $A_1$ and $A_2$, respectively. Also, $f_1$, $f_2$ are the cost functions for the respective local sub-problems. 
% \begin{equation}\label{eqProposed_obj1}
% \small
% \min_{X \in \mathcal{S}}{\hspace{0.2cm} F(X)}
% \end{equation}
\begin{equation}\label{eqProposed_obj2}
\small
\min_{X \in \mathcal{S}}{\hspace{0.1cm} F(X)} = \min_{X_1 \in \mathcal{S}_1, X_2 \in \mathcal{S}_2}{\hspace{0.1cm}} f_1(X_1,y_2)+f_2(X_2,y_1)
\end{equation}

The original problem defined in (\ref{eqProposed_obj2}) can be readily decomposable into the following two sub-problems (see \eqref{eqProposed_subprb1}), associated for respective decomposed areas; i.e., equation \eqref{sub1} and \eqref{sub2} for area $A_1$ and $A_2$, respectively.

\vspace{-0.3cm}
\begin{subequations}\label{eqProposed_subprb1}
\small
\begin{gather}
\text{For $A_1$}\hspace{0.3cm}: \min_{X_1 \in S_1}{\hspace{0.2cm}} f_1(X_1,y_2) \label{sub1}\\
\text{For $A_2$}\hspace{0.3cm}: \min_{X_2 \in S_2}{\hspace{0.2cm} f_2(X_2,y_1)} \label{sub2}
\end{gather}
\end{subequations}
\vspace{-0.3cm}

\noindent \textbf{\textit{Remark 1:}} Please note that, the decomposition of the OPFs also works for any maximization problem, such as \textbf{(C2)}.

\noindent \textbf{\textit{Remark 2:}} The decomposition method described here can easily be extended for a network, where multiple area decomposition is required to make each sub-problem small enough to make it solvable efficiently. Similar to the 2-area distributed OPF, the optimization problem can be decomposed into several smaller sub-problems, representing one decomposed area of the network. 

\noindent \textbf{\textit{Remark 3:}} The decomposition approach can be further extended to nodal decompositions, where each node represents one area.
\vspace{-3pt}
\subsection{Consensus for the Decomposed Sub-problems}
After decomposing the optimization problem into several smaller sub-problems, the proposed distributed algorithm solves the sub-problems individually to obtain respective local and complicating variables. Here, at each boundary among decomposed areas, the complicating variable $y_2$ and $y_1$ are kept fixed to solve sub-problem \eqref{sub1} and sub-problem \eqref{sub2}, respectively. Then the solved $y_1$ by sub-problem \eqref{sub1} and solved $y_2$ by sub-problem \eqref{sub2} are exchanged again between areas. After each macro-iteration, the update step of complicated variable, $\textbf{Y}$, is performed using Fixed Point Iteration (FPI) method, described by \eqref{fpi_up} for $n^{th}$ macro-iteration. Here, instead of a constant value, $\alpha$ can be made adaptive as well. The exchange process is repeated until the change in the all complicating variables for all decomposed boundaries over macro-iterations are within tolerance $\epsilon_{tol}$, see \eqref{tole}. 

\vspace{-0.2cm}
\begin{small}
\begin{equation}\label{fpi_up} 
\textbf{Y}^{(n)}:= \frac{\textbf{Y}^{(n)}+\alpha \textbf{Y}^{(n-1)}}{1+\alpha}
\end{equation}
\end{small}
\vspace{-0.6cm}

% \vspace{-0.2cm}
\begin{small}
\begin{equation}\label{tole} 
\Big|\textbf{Y}^{(n)} - \textbf{Y}^{(n-1)}\Big| \leq \epsilon_{tol}
\end{equation}
\end{small}
\vspace{-0.3cm}

In the next section, we will discuss the distributed approach to solve large scale OPF problems for radial power distribution networks using the decomposition method described in this section.

\section{Distributed OPF for Scalability}
In this section, we describe the distributed method to solve large scale OPF problems. For that, we use the decomposition technique, that we proposed previously. Specifically, we will detail the distributed algorithm to solve previously developed central OPFs, i.e., \textbf{(C1)}, \textbf{(C2)}, and \textbf{(C3)}. First, we discuss the formulation of the sub-problems, and then we describe the algorithm.
\vspace{-2pt}
\subsection{Distributed Sub-Problems}
For a system decomposed into multiple areas, the sub-problems are defined for each $A_m \in \A_R$, where $A_m = \G(\N_m, \E_m)$. Please note, while decomposing the network, it is ensured that the number of nodes/variables in each decomposed areas do not exceeds a certain number, that might cause computation complexities. Here, the power flow model is defined in \eqref{Local_PF1}-\eqref{Local_PF3}, and are used by the corresponding sub-problem for area $A_m$ -- defined $\forall j \in \N_m$ and $\forall \{ij\},\{jk\} \in \E_m$. Also, we let $C_h$ be the set of buses, that is shared by area $A_m$ with its child areas. The sub-problems for (i) loss minimization, (ii) DER maximization, and (iii) $\Delta$V minimization is detailed next. For these OPF objectives, we use the same OPF formulation as central problem, except only for the corresponding area, $A_m$. Also, the shared bus voltage and power flow (complicated variables) are updated using \eqref{fpi_up}, and kept fixed for $n^{th}$ iteration, as shown in equation \eqref{Local_PF4}-\eqref{Local_PF6}. The sub-problem for loss minimization is described below, \eqref{lm_local}.

\vspace{-0.3cm}
\begin{small}
\begin{IEEEeqnarray}{C C}
\IEEEyesnumber\label{lm_local} \IEEEyessubnumber*
\text{\textbf{(D1)}}\hspace{0.4cm} \min \hspace{0.2cm} f_m = \sum_{\{ij\}\in \E_m} l_{ij}r_{ij} \hspace{0.6cm}\\
\text{s.t.}  \hspace{0.2cm} P_{ij}-r_{ij}l_{ij}-p_{L_j}+p_{Dj}= \sum_{k:j \rightarrow k} P_{jk} \label{Local_PF1} \\
Q_{ij}-x_{ij}l_{ij}-q_{L_j}+q_{Dj}= \sum_{k:j \rightarrow k} Q_{jk}  \label{Local_PF2} \\
v_j=v_i-2(r_{ij}P_{ij}+x_{ij}Q_{ij})+(r_{ij}^2+x_{ij}^2)l_{ij}  \label{Local_PF3}\\
\vspace{1pt}
v_{o} = v_{o'} ; \label{Local_PF4}\\
\vspace{1pt}
P_{jk} = p_{k'} \hspace{0.2cm} ; \forall \{jk\} \in \E_m \text{, where } k \in C_h \label{Local_PF5}\\
\vspace{1pt}
Q_{jk} = q_{k'} \hspace{0.2cm} ; \forall \{jk\} \in \E_m \text{, where } k \in C_h \label{Local_PF6}\\
\vspace{2pt}
-\sqrt{S_{DRj}^2-p_{Dj}^2} \leq q_{Dj} \leq \sqrt{S_{DRj}^2-p_{Dj}^2}\\
\underline{V}^2 \leq v_j \leq \overline{V}^2 \hspace{0.7cm} ;\forall j\in \N_m \label{lm_local2}\\
l_{ij} \leq \left(I^{rated}_{ij}\right)^2  \hspace{0.5cm} ;\forall \{ij\} \in \E_m \label{lm_local3}
\end{IEEEeqnarray}
\end{small}
\vspace{-3pt}

\noindent Here, area $A_m$ shares bus '$o$' with its parent area, and $v_{o'}$ is the solved bus voltage by that parent area in the previous iteration. Similarly, $\forall k\in C_h$,  $,~p_{k'},\&~ q_{k'}$ is the solved shared power flows by the child areas of $A_m$ in the previous iteration. Note that, the symbol $|.|'$ depicts that the variable is solved by other areas, and not by the area that is associated with the corresponding sub-problem. Further, the sub-problems for DER maximization and $\Delta$V minimization is formulated in \eqref{dm_local} and \eqref{dv_local}, respectively.

% \vspace{-0.6cm}
\begin{small}
\begin{IEEEeqnarray}{C C}
\IEEEyesnumber\label{dm_local} \IEEEyessubnumber*
\text{\textbf{(D2)}}\hspace{0.4cm} \max \hspace{0.2cm} f_m = \sum_{j\in \N_m} p_{D_j} \hspace{0.6cm}\\
\text{s.t.}  \hspace{0.2cm} \text{\eqref{Local_PF1} - \eqref{Local_PF6}, \eqref{DG_lim2},  \eqref{lm_local2} - \eqref{lm_local3}}
\end{IEEEeqnarray}
\end{small}
% \vspace{-1cm}
\begin{small}
\begin{IEEEeqnarray}{C C}
\IEEEyesnumber\label{dv_local} \IEEEyessubnumber*
\text{\textbf{(D3)}}\hspace{0.4cm} \min \hspace{0.2cm} f_m = \sqrt{\sum_{\forall j\in \N_m}(v_j-v_{ref})^2} \hspace{0.6cm}\\
\text{s.t.}  \hspace{0.2cm} \text{\eqref{Local_PF1} - \eqref{Local_PF6}, \eqref{DG_lim},  \eqref{lm_local2} - \eqref{lm_local3}}
\end{IEEEeqnarray}
\vspace{-3pt}
\end{small}

\vspace{-12pt}
\subsection{Algorithm}
For completeness, now we discuss the full distributed algorithm that decomposes the OPFs for large scale power distribution systems, and solves iteratively to reach the global solution. Here, we use the decomposition technique that we developed in Section III, and solve sub-problems for different network level objectives, i.e., \textbf{(D1)}-\textbf{(D3)} until convergence. We use the same notation for network variables that we denoted for decomposed areas in Section IIIA and IIIB. We use tolerance of $\epsilon_{tol} = 0.001$ to meet the convergence criterion. The algorithm is detailed below (see Algorithm 1). To better understand the distributed computing of the OPFs, the sub-routine in step 5 of Algorithm 1 is described in Algorithm \ref{stp4},
%%%%%%%%%%%%%%%%%%%%%%%%%%%%%%%%%%%%%%%%%%%%%%%%%%%%
\vspace{-8pt}
\begin{algorithm}[t] 
    \small
    \caption{\small Distributed Algorithm for Scaled OPFs}
    \SetKwInOut{Stp}{Steps}
    Decompose the network into $N$ areas, so that each area has a maximum specified node numbers\\
    {Initialize complicating variables, ${\textbf{Y}}^0 \in \mathcal{S}$; error, $e = 1$; and macro-iteration count $n = 0$}\\
    If {$|e| \leq \epsilon_{tol}$,} stop iteration count, and go to step 10\\
    \vspace{3pt}
    Else, increase iteration count $n$: $n \leftarrow n+1$\\
    Solve $\Phi_m$ in parallel using \textbf{Algorithm 2}, for all decomposed areas, where, $\Phi_m$ depicts the solution of sub-problems;
    $\Phi_m:\hspace{0.1cm}X_m^{(n)}:= \underset{X_m \in S_m}{\text{argmin/argmax}}\hspace{0.2 cm} f_m\Big(X_m, {y}_{m^'}^{(n-1)}\Big)$\\
        \vspace{3pt}
    Update all the complicating variables, $\textbf{Y}$, using \eqref{fpi_up}, where $\alpha$ can be constant or adaptive\\
    \vspace{3pt}
    {Check residual vector}
    $\mathcal{R}^{(n)}=
    \left[
    \begin{array}{c}
    \textbf{Y}^{(n)} - \textbf{Y}^{(n-1)}
    \end{array}
    \right]$\\
    \vspace{3pt}
    $e = \max \hspace{3pt} |\mathcal{R}^{(n)}|$\\
    Go to step 2\\
        \vspace{3pt}
{Return Global Minimizer: $X^* = \{X_m^n ~|~ m=1,2,...,N\}$}
%   \noindent\rule{8.2cm}{1pt}
    \label{algori1}
        % \vspace{-1.1cm}
\end{algorithm}
\vspace{-8pt}

\begin{algorithm}[t]
\small
\caption{\centering{\small Sub-routine to Solve Sub-Problems at Step-5}}\label{stp4}
	%\noindent\rule{8.4cm}{1pt}\\ 
\SetAlgoLined
\SetKwInOut{Rx}{Complicated Variables }
\SetKwInOut{Tx}{Optimization Variable }
\SetKwInOut{St}{Calculate}
\SetKwInOut{ND}{Sub-Problem}
\SetKwInOut{TE}{Macro-Iteration step}
\SetKwInOut{Stp}{Steps }
\ND {For decomposed area $A_m \in \A_R$}
\TE{n}
% \St{q^*_{Dj}}\\
\Rx{$y_{m^'}^{(n-1)}$, variables that are used for coupling sub-problems}
\vspace{3pt}
\Tx{$X_m$}
\vspace{3pt}
\Stp{ }
Assume the complicating variable for area $A_m$ as constant; i.e., $y_{m^'}^{(n-1)}$ is set to either constant voltage (if it has parent area), or constant loads (if it has child areas), or both -- depending on the position of the area $A_m$ (See equation \eqref{Local_PF4}-\eqref{Local_PF6})\\
\vspace{2pt}
Solve distributed sub-problems of minimizing or maximizing the decomposed cost function $f_m$, e.g., \textbf{(D1)}, \textbf{(D2)}, etc. by assuming the constant complicating variables by off-the-shelf NLP solvers\\
Store the local minimizer in the variable $X_m^{(n)}$
	\label{algo}
% 	\vspace{-0.2cm}
\end{algorithm}
% \vspace{-3pt}

%%%%  Numerical Simulations and Results section: %%%%%%%%%
% \vspace{-0.3cm}
\section{Numerical Simulation}
\vspace{-3pt}
To show the efficacy and validate the proposed decomposition approach to solve the large scale OPFs for power distribution systems, we simulate our algorithm for a very large scaled, balanced synthetic 10,000 node distribution system and medium-scale balanced IEEE-8500 node test system with 2500 nodes. All experiments were simulated in Matlab 2018b on a machine with 8GB memory and Core i7-8700 CPU @3.19 GHz. The NLP sub-problems of the distributed method is solved using \textit{fmincon} of Matlab using \textit{'sqp'} algorithm. However, given the NP-hard nature of the centralized OPF problems, for bench-marking against centralized OPF, we use a commercial NLP solver \textit{Artelys Knitro} with \textit{'active-set'} algorithm, that scales relatively well with the problem size  \cite{nocedal2006knitro}.
\vspace{-3pt}
\subsection{Simulated System}
\begin{figure}[t]
    \centering
    \includegraphics[width=0.48\textwidth]{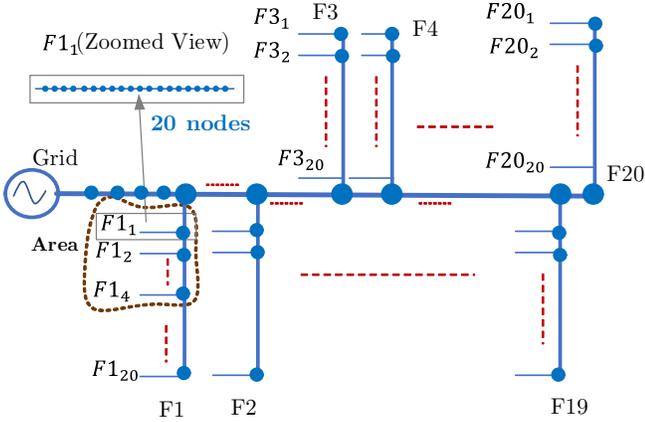}
        \vspace{-0.2cm}
    \caption{Synthetic 10,000 Node System}
    \label{syn_10k}
% \vspace{-0.5cm}
\end{figure}

\vspace{-3pt}
The simulations are conducted using the following two test systems: (1) Synthetic 10,000 node distribution system with different DER penetration levels, and (2) Balanced IEEE-8500 node test system with 100\% DER penetration for nodal decomposition. Please note, the \% penetration means the percentage of DER nodes compare to load nodes. The Synthetic 10,000 node system is shown in Fig. \ref{syn_10k}. The distribution system is comprised of 1 main-feeder, 20 laterals, where each lateral supplies 20 neighborhoods. It is assumed that each neighborhood is comprised of 20 households. Thus, each lateral supplies a total of 400 houses. Also, in between 2 laterals, we assume 4 nodes in the main-feeder that represent the distributed loads. Every load in this distribution network is set to consume a total of $S_L = 0.1+0.01j$ pu, and the line impedance of all the branches is assumed to be $z = 0.07+0.01j$ pu. The base voltage for the network is 12.47 kV ($V_{LL}$) and base kVA is 1000. For loss minimization and $\Delta$V minimization objectives, each DER in the system can generate 7 kW of real power, with nominal rating of 8.4 kVA. For DER maximization problem, the rating of the DERs are increased to at-most 10 times to stress the system.
% , and further validate the efficiency of the proposed method. 
We decompose the distribution system in multiple areas where each area is composed of 100 nodes (see Fig. 2).  

For the IEEE 8500 node test system, the DER sizes are chosen randomly with a rating ranging from 1.3 to 5.8 kVA. We use this medium-scale distribution system to further decompose the problem into nodal level, i.e., each node is considered as an area. Please note that, this is a balanced, single-phase equivalent distribution network of the test system, that has 2522 nodes. We used the same base values, i.e., 12.47 kV as base voltage and 1000 kVA as base kVA for this test system. The proposed decomposition technique is next simulated for various DER penetration with different network objectives to solve the OPFs for these scaled-networks. 

\subsection{Loss Minimization Objective: (D1)} First, we solve the loss minimization problem \textbf{(D1)} for 10,000 node test system with varying DER penetration levels. The active power loss in the network is minimized by the reactive power generations of the DERs. For loss minimization OPFs, we have used the nominal case for synthetic 10,000 node system. That is  all the loads are at nominal value ($S_L$).
% , and DERs are capable of generating 7 kW of active power, i.e., 0.07 pu.
The kVA rating of these DERs are 120\% of the nominal active power generation. We have simulated (i) 100\%, (ii) 50\%, and (iii) 10\% DER penetration levels for loss minimization objective with grid voltage of 1.00 pu. The result of this OPF is detailed next.
% The sub-problems \textbf{(D1)} is solved by \textit{fmincon} with \textit{sqp} algorithms.
We have used $\alpha = 0$ for FPI update in \eqref{fpi_up}.

\begin{figure*}[t]
\centering
\begin{subfigure}{.24\textwidth}
  \centering
  \includegraphics[width=1.05\linewidth]{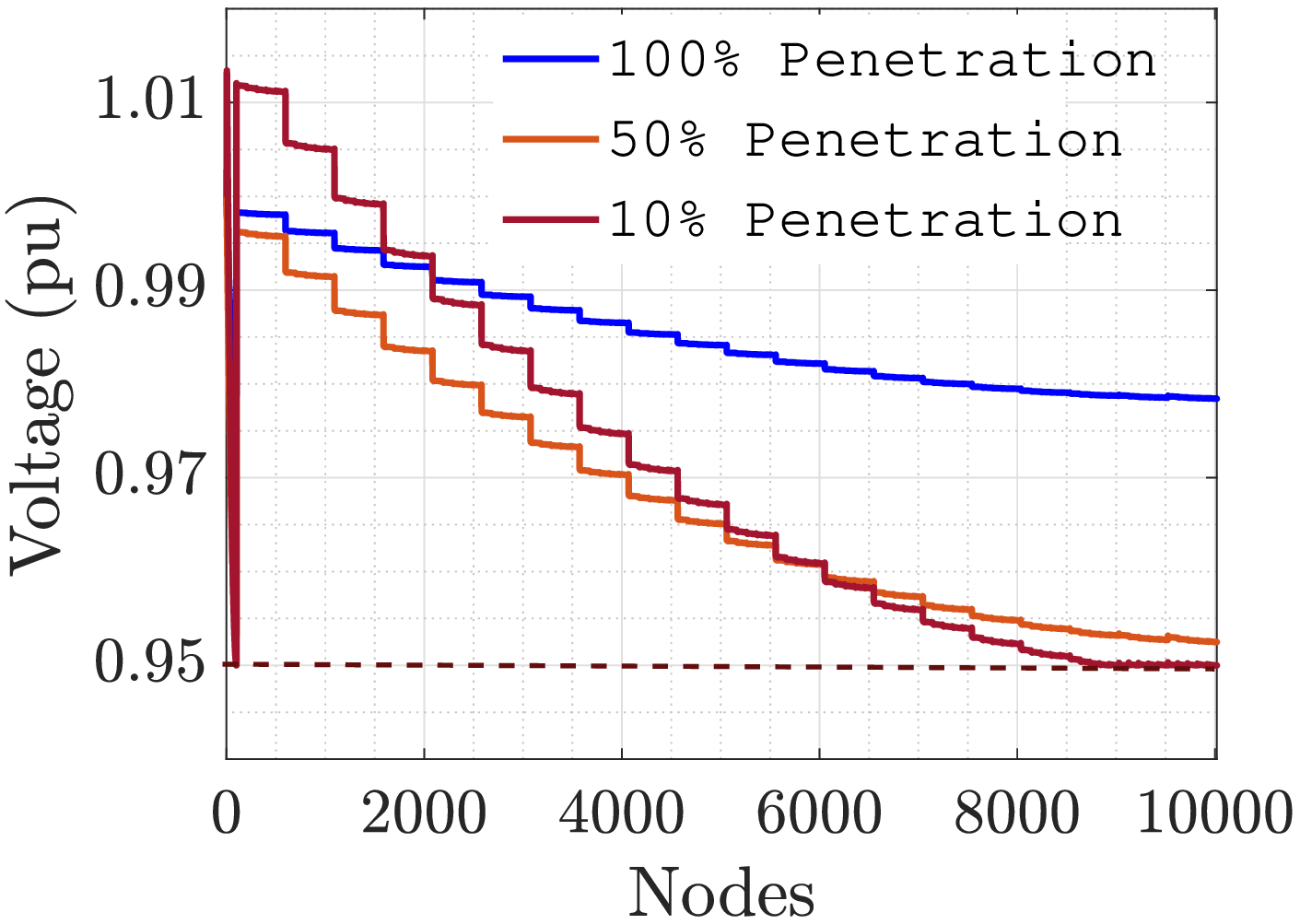}
  \caption{Nodal voltages}
  \label{volt_lm}
\end{subfigure}
\begin{subfigure}{.24\textwidth}
  \centering
  \includegraphics[width=1.05\linewidth]{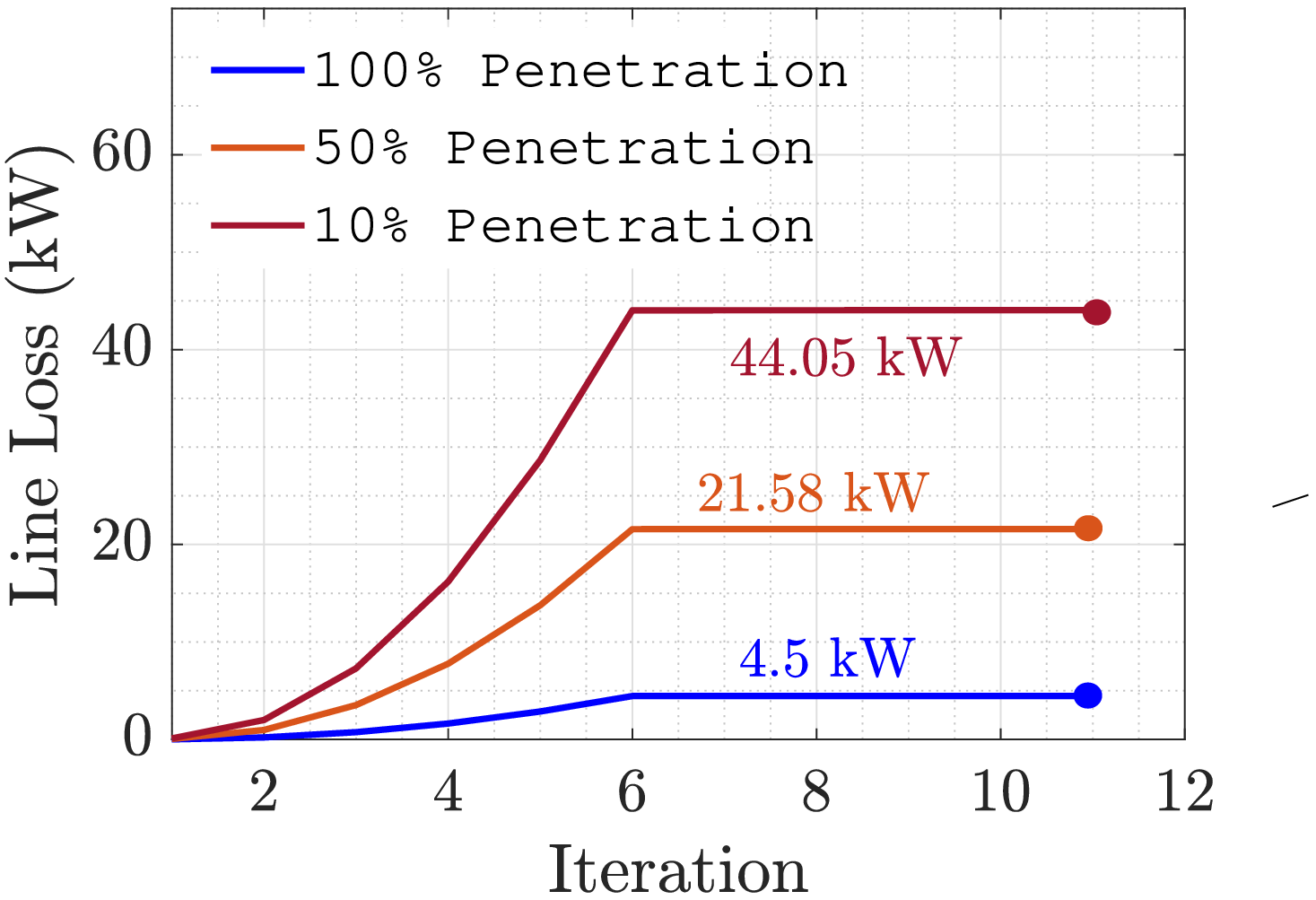}
  \caption{Objective value}
  \label{obj_lm}
\end{subfigure}
\begin{subfigure}{.24\textwidth}
  \centering
  \includegraphics[width=1.05\linewidth]{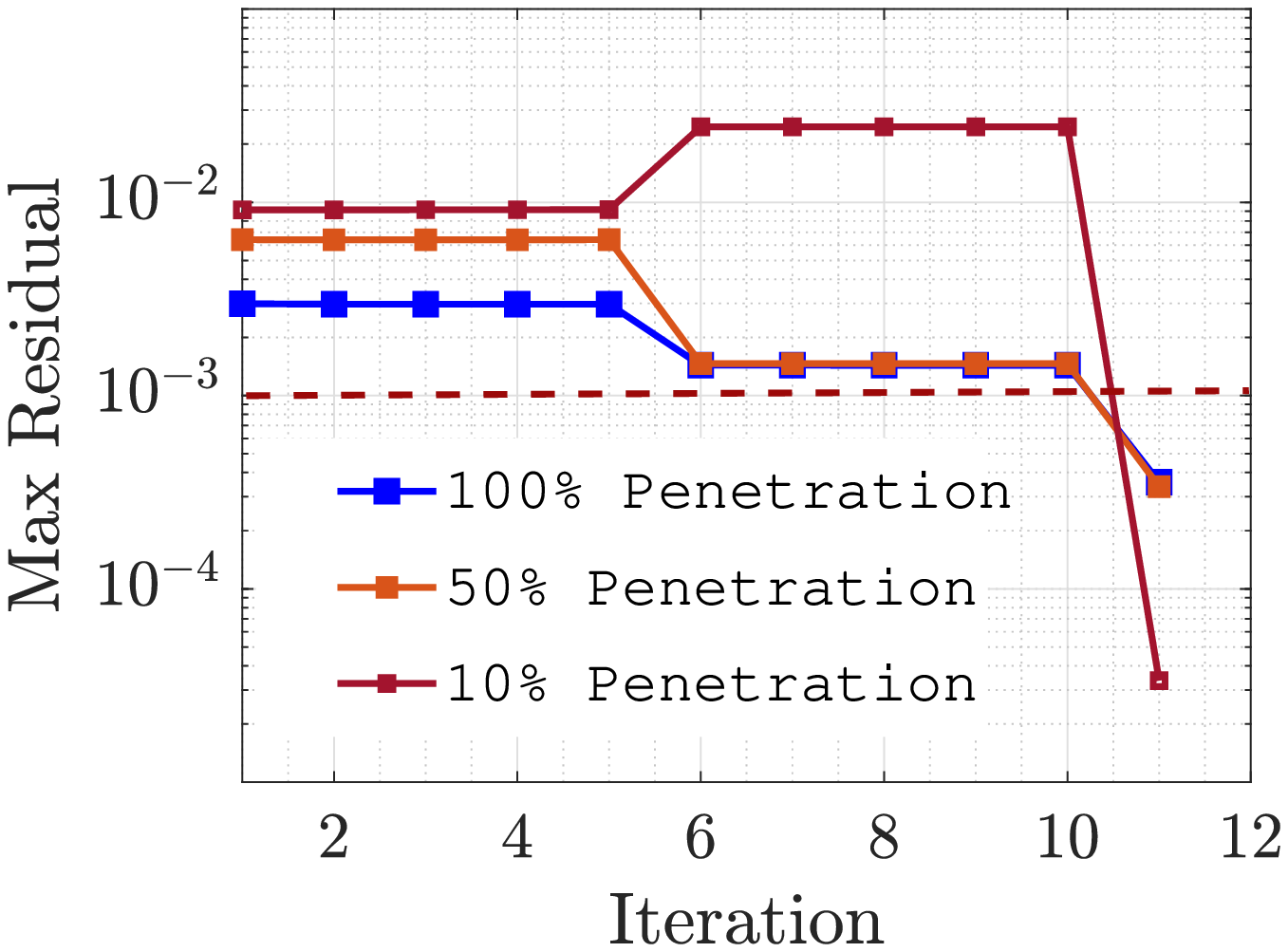}
  \caption{Convergence at the boundary}
  \label{R_lm}
\end{subfigure}
% \hspace{-1 pt}
\begin{subfigure}{.24\textwidth}
  \centering
  \includegraphics[width=1.05\linewidth]{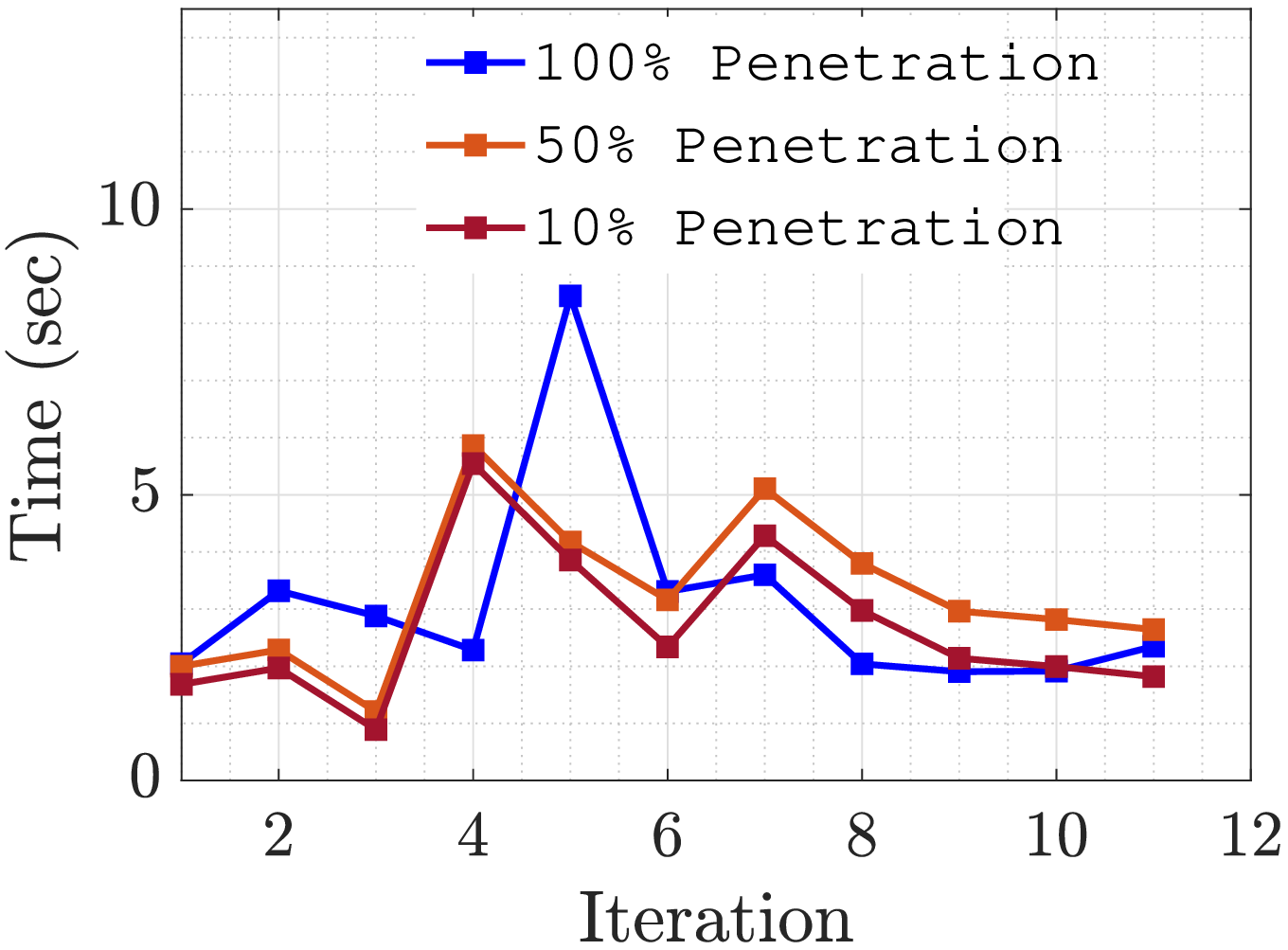}
  \caption{ \centering{Simulation time}}
  \label{t_lm}
\end{subfigure}
\caption{Numerical Results for Loss Minimization Objective for Synthetic 10,000 Node System}
\label{Result_lm}
\end{figure*}

% \subsubsection{Convergence for \textbf{D1}} 
The converged solution of the decomposed central OPF and the convergence properties of the  proposed method for the loss minimization problem is shown in Fig. \ref{Result_lm}. We can see that the converged voltage does not violate any voltage constraints, i.e., the voltage is not outside of the specified limit of 0.95-1.05 pu bound (Fig. \ref{volt_lm}). Also, with increased penetration, the nodal voltages over the network has less standard deviation (S.D.), that results in lower active power loss in the network. The nodal voltages has higher S.D. for 10\% DER penetration, i.e., more spreaded for less DER penetration. Fig. \ref{obj_lm} shows the objective value of the OPF problem w.r.t. macro-iterations. 100\% penetration can reduce the line losses to 4.5 kW, but with lower DER penetrations, the line losses increases. The convergence properties for this case is shown in Fig. \ref{R_lm}. For all the cases, it takes around 11 macro-iterations to meet the convergence criterion. Besides, the time taken at each iteration for this case is plotted in Fig. \ref{t_lm}. This time represents the highest time taken to solve any sub-problem at each iteration. It only takes $\sim 30$ seconds to solve the OPF by decomposing the problems into several sub-problems for all the DER penetration levels (see Table \ref{res_tab}). 

\begin{figure*}[t]
\centering
\begin{subfigure}{.24\textwidth}
  \centering
  \includegraphics[width=1.05\linewidth]{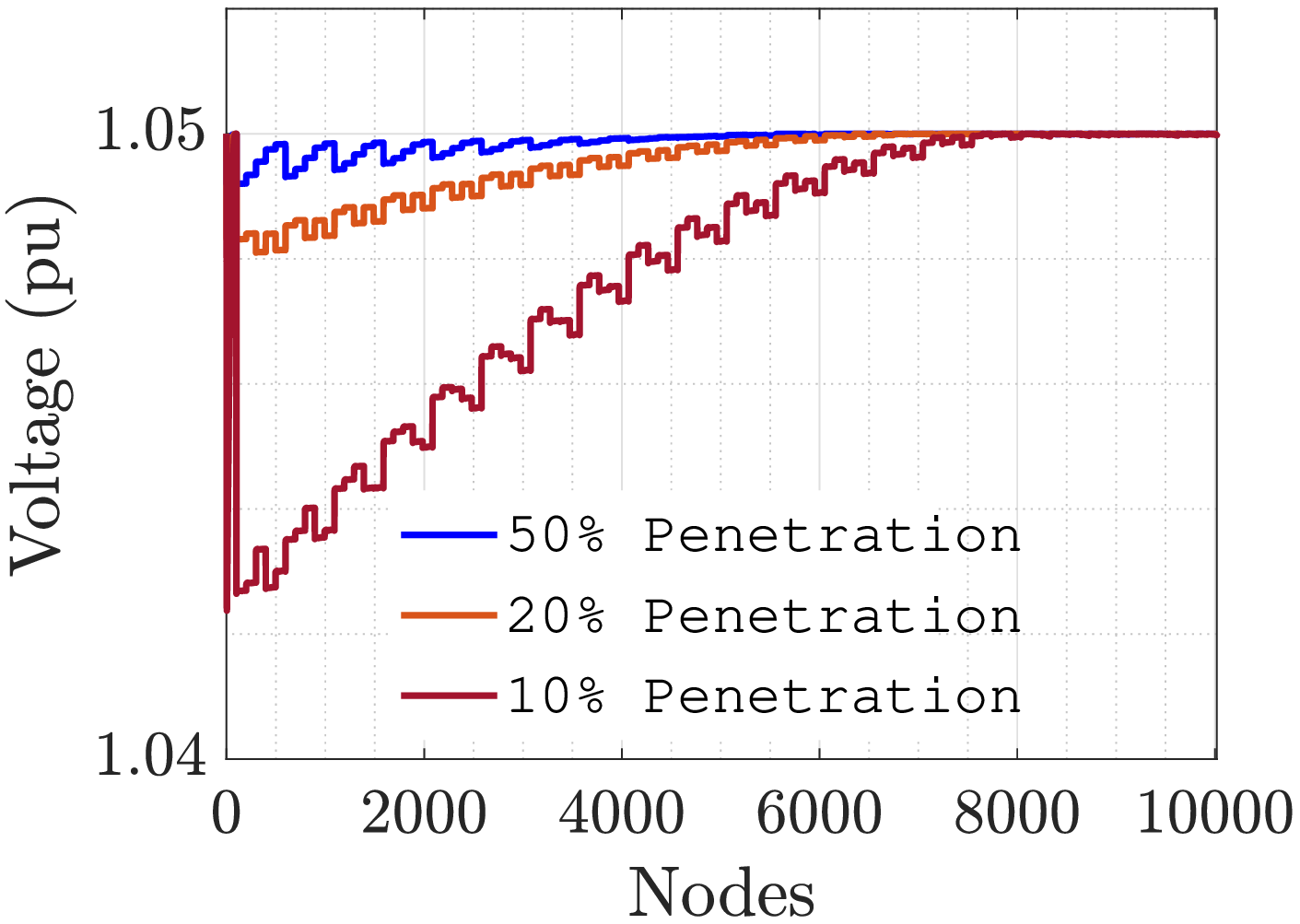}
  \caption{Nodal voltages}
  \label{volt_pm}
\end{subfigure}
\begin{subfigure}{.24\textwidth}
  \centering
  \includegraphics[width=1.05\linewidth]{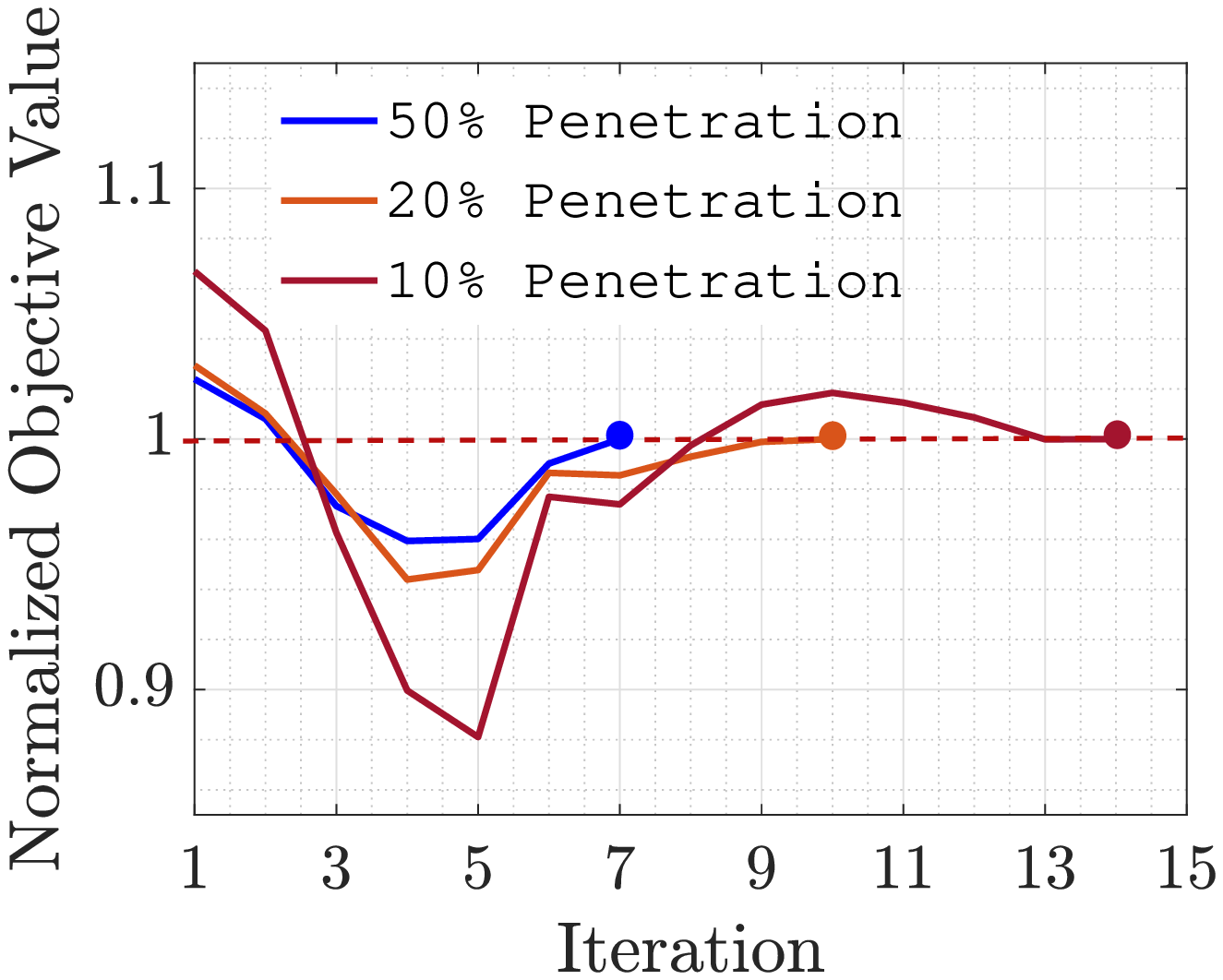}
  \caption{Objective value}
  \label{obj_pm}
\end{subfigure}
\begin{subfigure}{.24\textwidth}
  \centering
  \includegraphics[width=1.05\linewidth]{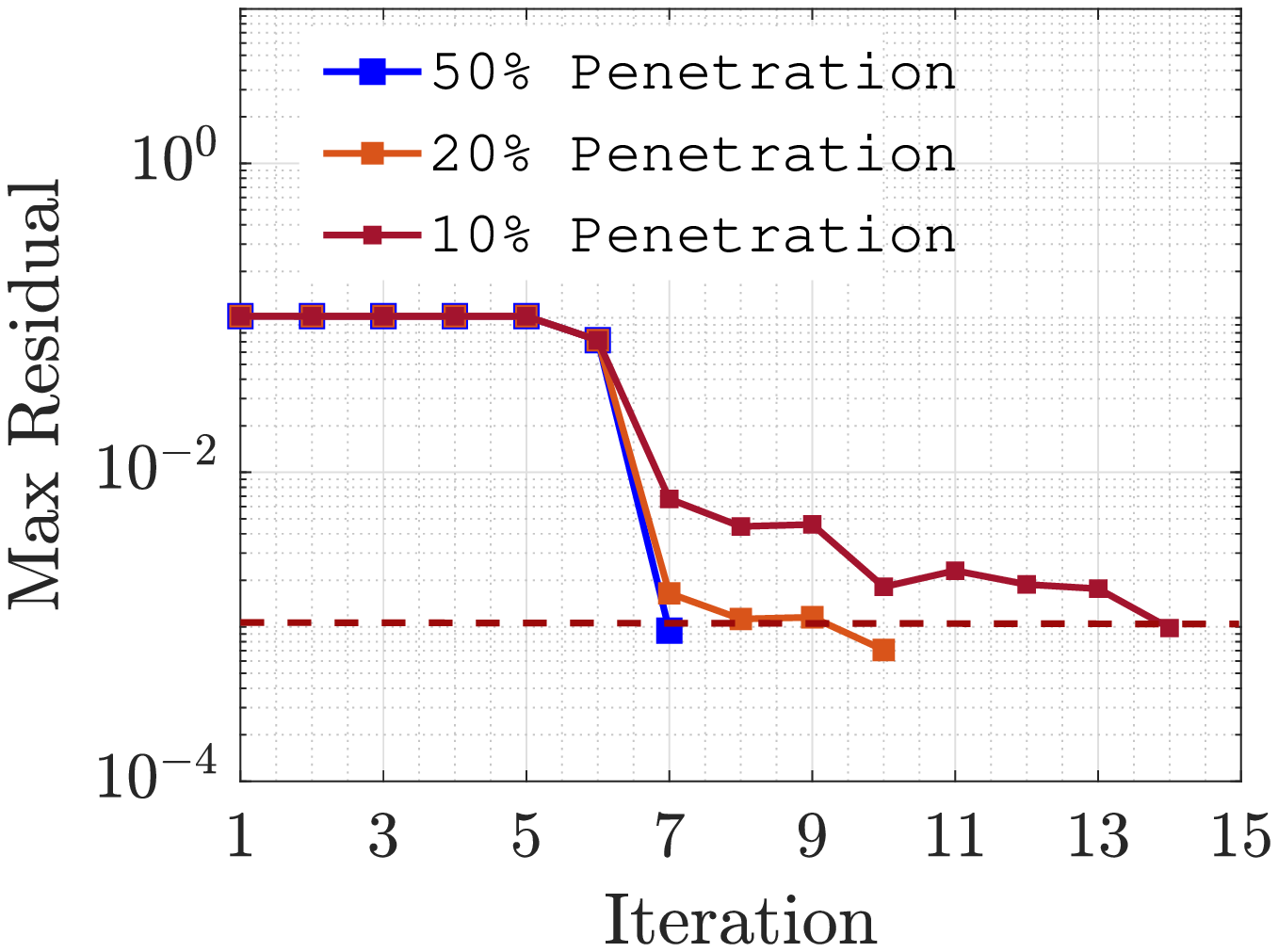}
  \caption{Convergence at the boundary}
  \label{R_pm}
\end{subfigure}
% \hspace{-1 pt}
\begin{subfigure}{.24\textwidth}
  \centering
  \includegraphics[width=1.05\linewidth]{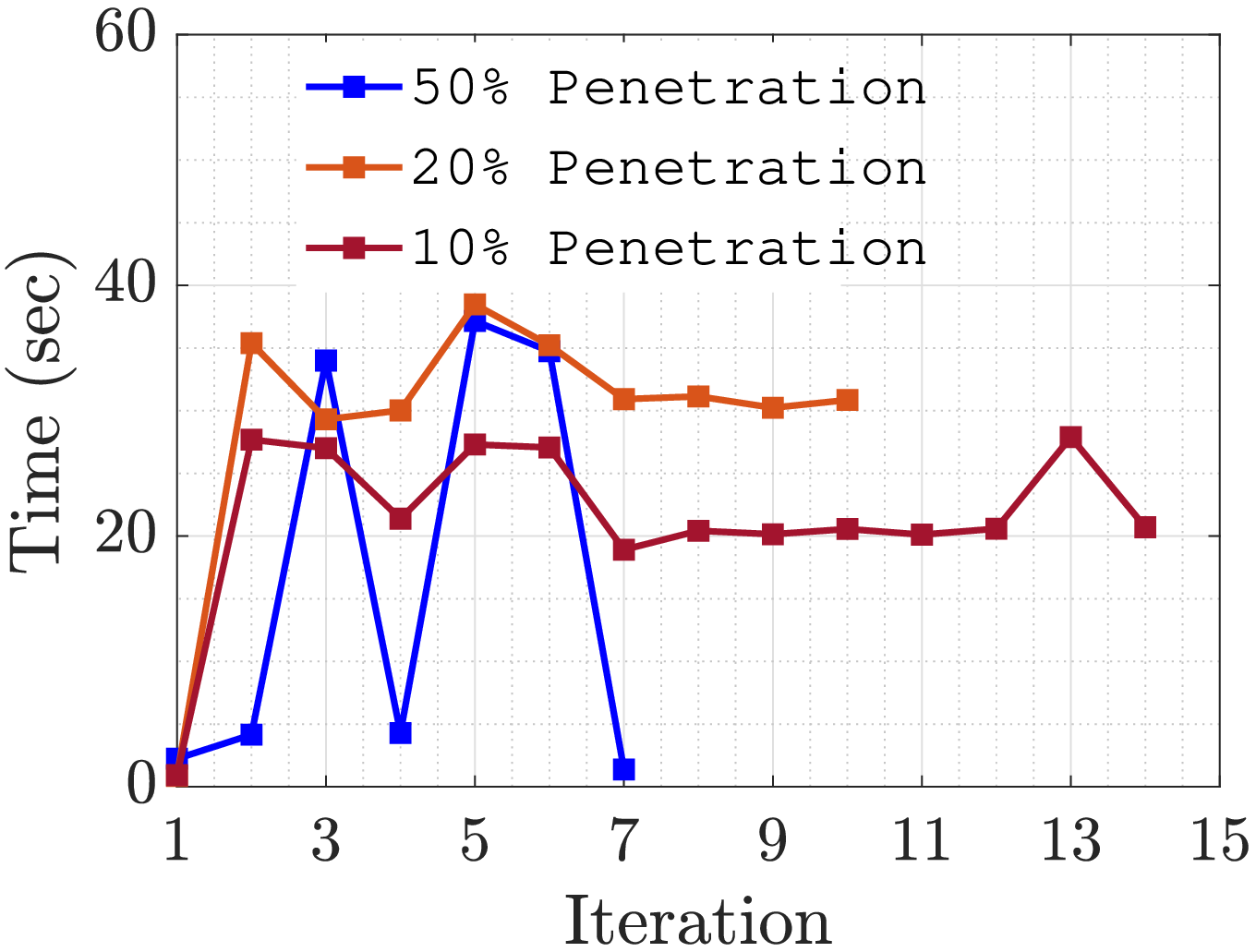}
  \caption{ \centering{Simulation time}}
  \label{t_pm}
\end{subfigure}
\caption{Numerical Results for DER Maximization Objective for Synthetic 10,000 Node System}
\label{Result_pm}
\end{figure*}

\subsection{DER Maximization Objective: (D2)}
In this section, we present the result for DER maximization OPF problem, which is also equivalent to DER curtailment problem for the power distribution networks. Here, similar to the previous OPF problem, we solve the decomposed problem \textbf{(D2)} for 10,000 node test system with different DER penetration levels. The active power generation of the DERs are maximized while maintaining the operation limits of the network, such as voltage limits. For this optimization problem, we have used various load and generation multiplier to stress the system at max level. We simulate 3 different cases -- (i) 50\% DER penetrations where the active power generation capacities of DERs are 21 kW and loads are nominal, (ii) 20\% DER penetrations with 28 kW of active power generation capacities for each DERs and load multiplier is 0.5, and (iii) 10\% DER penetration levels with max of 70 kW generation capabilities and load multiplier is 0.5. The grid is assumed to be operating at 1.05 pu.
% We have used the same NLP solver (fmincon with sqp) for this case as well, and 
The result of this OPF is discussed next. We have used $\alpha = 2.33$ for FPI update in \eqref{fpi_up}.

% \subsubsection{Convergence for \textbf{D1}} 
From the Fig. \ref{Result_pm}, we can see the converged solution of DER maximization OPF, that has been solved using proposed decomposition method. Similar to the previous objective, 
we can see that the nodal voltage does not violate the voltage limits (Fig. \ref{volt_pm}). The voltages are near the upper bound shows that the systems were highly stressed for these different simulated cases. With increased penetrations, more nodes have voltages that are closer to the upper limits of 1.05 pu. Fig. \ref{obj_pm} shows the normalized objective value of the OPF problem w.r.t. macro-iterations. Here, the objective value is scaled w.r.t. the converged/final cost as the orders of the final costs are different. The actual values of the objective function upon solving OPFs using distributed approach is shown in table \ref{res_tab}. Note that, even though the number of DERs in 10\% penetration case is lower than 20\% case, but individual DERs has higher capacity in 10\% penetration case than later, and thus total generation is higher in 10\% penetration case than 20\% penetration case. It takes 7, 10 and 14 iterations to converge for 50, 20 and 10\% DER penetrations (Fig. \ref{obj_pm},\ref{R_pm} ). The simulation time per macro-iteration is shown in Fig. \ref{t_pm}. The total simulation time to solve the OPF using proposed decomposition approach is reported in Table \ref{res_tab}.

\begin{small}
\begin{table}
    % \vspace{-0.3cm}
    \centering
    \caption{Results Summary}
    \vspace{-0.1cm}
    \label{res_tab}
    \setlength{\tabcolsep}{3.0pt}
    \begin{tabular}{|c|c|c|c|}
    \hline
    
    \multirow{2}{*}{\textbf{OPF Problem}}  & \multirow{2}{*}{\textbf{DER\%}} & \multirow{2}{*}{\textbf{Converged Objective Value}}  & \multirow{2}{*}{\textbf{Time (s)}}\\
&&&\\
    \hline
    \multirow{3}{*}{Loss Min} & 100 & 4.5 kW & 34\\
        & 50 & 21.58 kW & 35\\
        & 10 & 44.05 kW & 30\\
    \hline
    \multirow{3}{*}{DER Max} & 50 & 1.03 MW (Total capacity 1.05 MW) & 120\\
        & 20 & 0.55 MW (Total capacity 0.56 MW) & 240\\
        & 10 & 0.66 MW (Total capacity 0.70 MW) & 300\\
    \hline
    \multirow{2}{*}{$\Delta$V Min} & 100 & 2.65 pu & 30\\
        & 50 & 6.68 pu & 15\\
    \hline
    % \vspace{-0.8cm}
    \end{tabular}
\end{table}
\end{small}

\subsection{$\Delta$V Minimization Objective: (D3)}
Now we show the result for the third objective function, i.e., $\Delta$V minimization problem. Here, we solve the decomposed problem \textbf{(D3)} for 10,000 node test system with different DER penetration levels, but with nominal values of DER generation and loads. The reactive power generation of the DERs are controlled to make the nodal voltages closer to a reference value of $V_{ref} = 1.00 pu$, which is the substation node voltage. For this optimization problem, simulated 2 different cases -- (i) 100\% DER penetrations and (ii) 50\% DER penetrations. Again, we set $\alpha = 0$ for FPI update.
% , and used the same NLP solver (fmincon with sqp) for this case.

\begin{figure}[t]
\centering
\begin{subfigure}{.24\textwidth}
  \centering
  \includegraphics[width=1.00\linewidth]{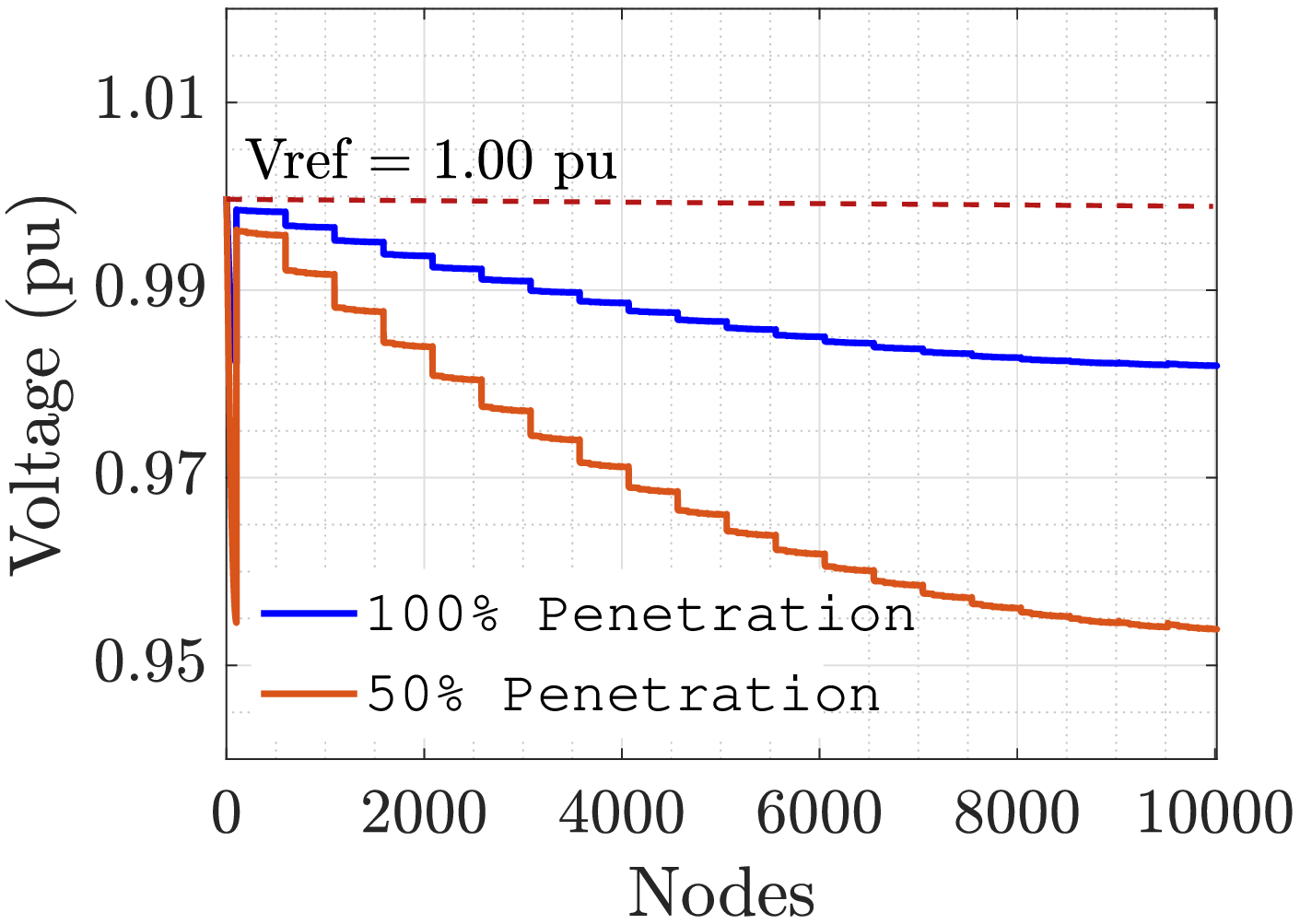}
  \caption{Nodal voltages}
  \label{volt_vm}
\end{subfigure}
\begin{subfigure}{.24\textwidth}
  \centering
  \includegraphics[width=1.0\linewidth]{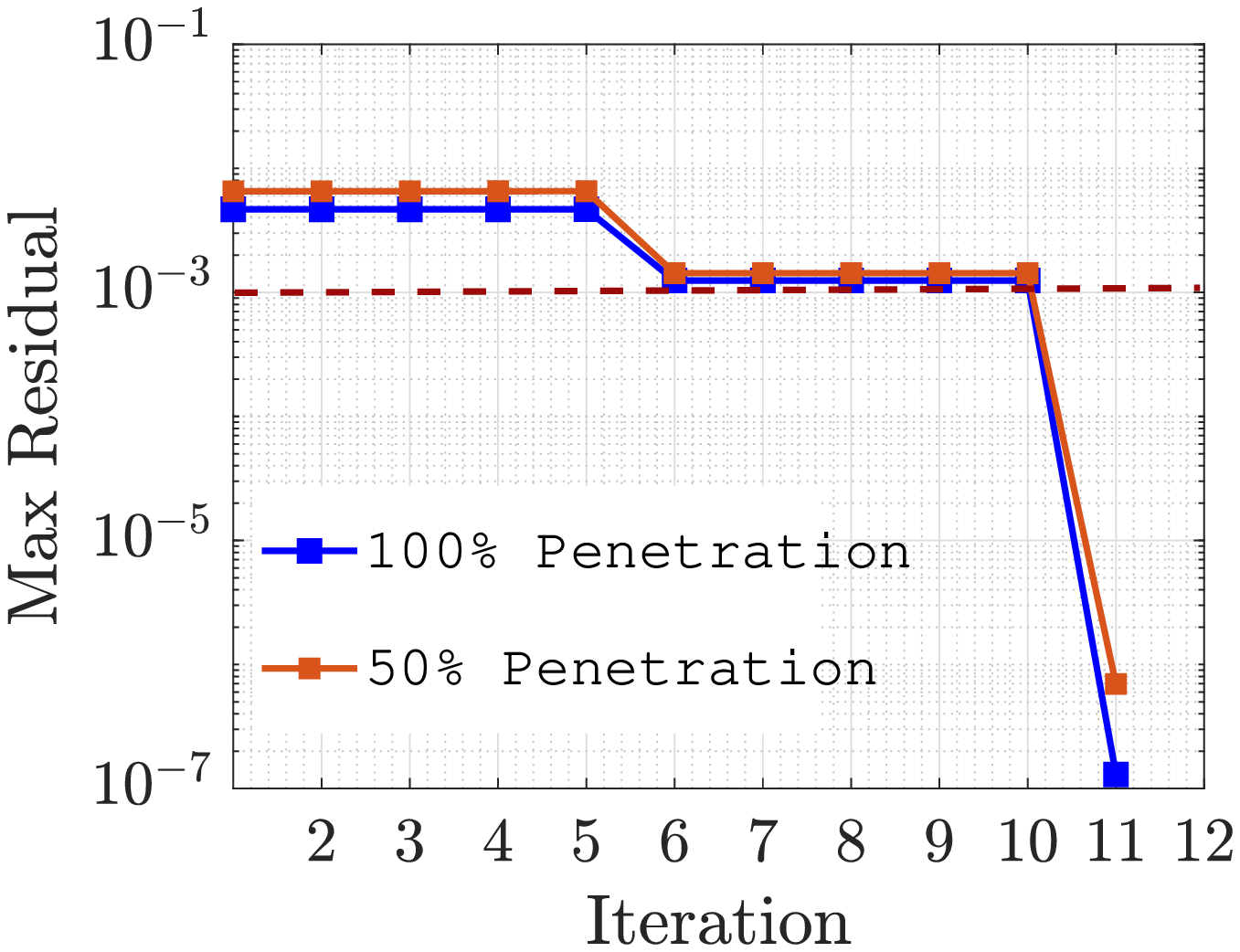}
  \caption{Convergence at the boundary}
  \label{R_vm}
\end{subfigure}
\caption{\centering{Numerical Results for $\Delta$V Minimization Objective for Synthetic 10,000 Node System}}
\label{Result_vm}
\end{figure}

% \subsubsection{Convergence for \textbf{D1}}
The optimal result is shown in Fig. \ref{Result_vm}, where Fig. \ref{volt_vm} shows that the nodal voltages after optimization. The higher penetrations of DERs results in closer node voltages to the substation voltage. Also, for both of the cases, it only takes 11 macro-iterations to reach convergence (Fig. \ref{R_vm}). The objective value of this cost function is shown in Table \ref{res_tab} with the total solution time. The time taken at each iteration is almost same for all the iterations, and thus not shown in this paper.

\begin{figure*}[t]
\centering
\begin{subfigure}{.32\textwidth}
  \centering
  \includegraphics[width=1.00\linewidth]{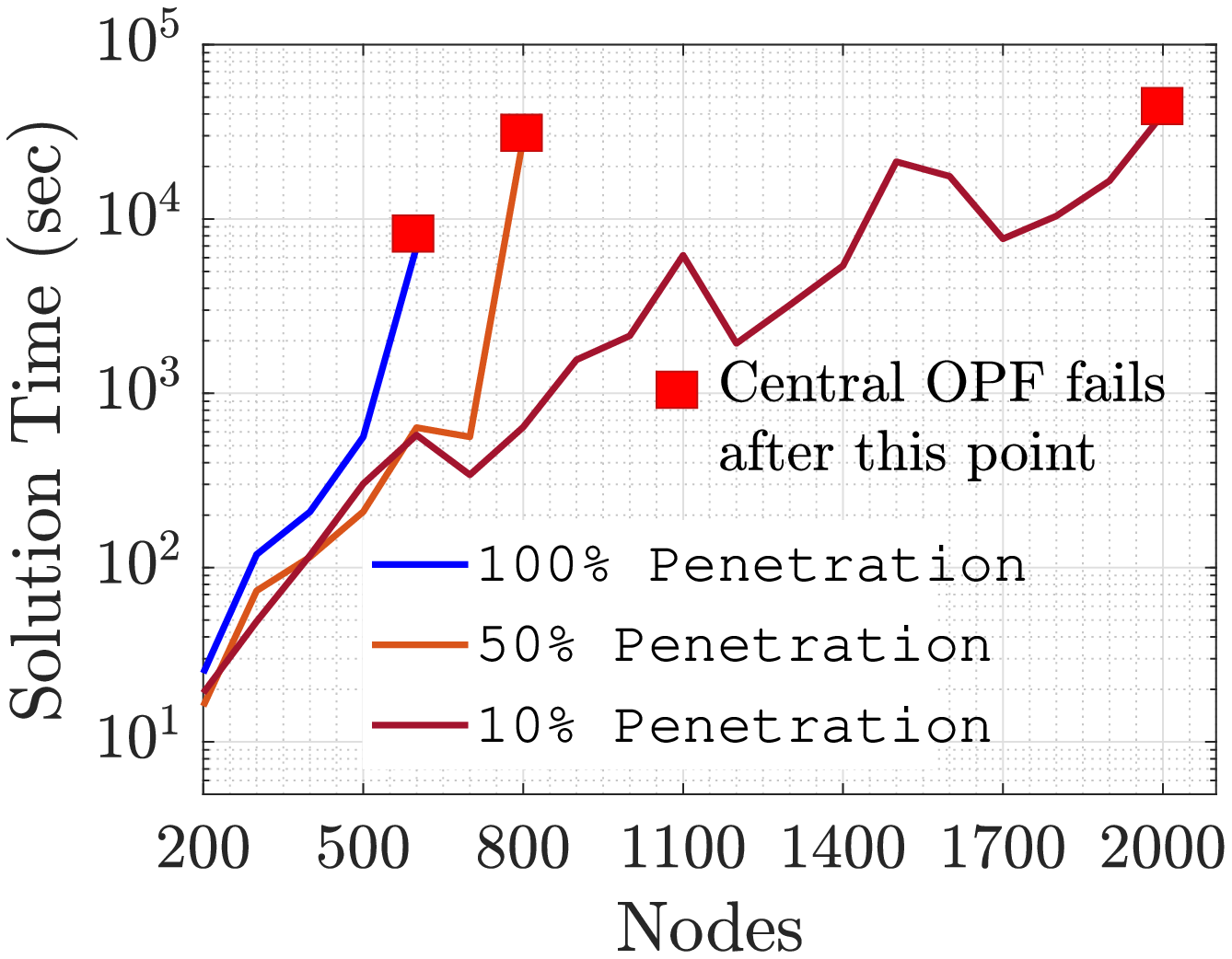}
  \caption{Loss minimization objective}
  \label{fail_lm}
\end{subfigure}
\begin{subfigure}{.32\textwidth}
  \centering
  \includegraphics[width=1.00\linewidth]{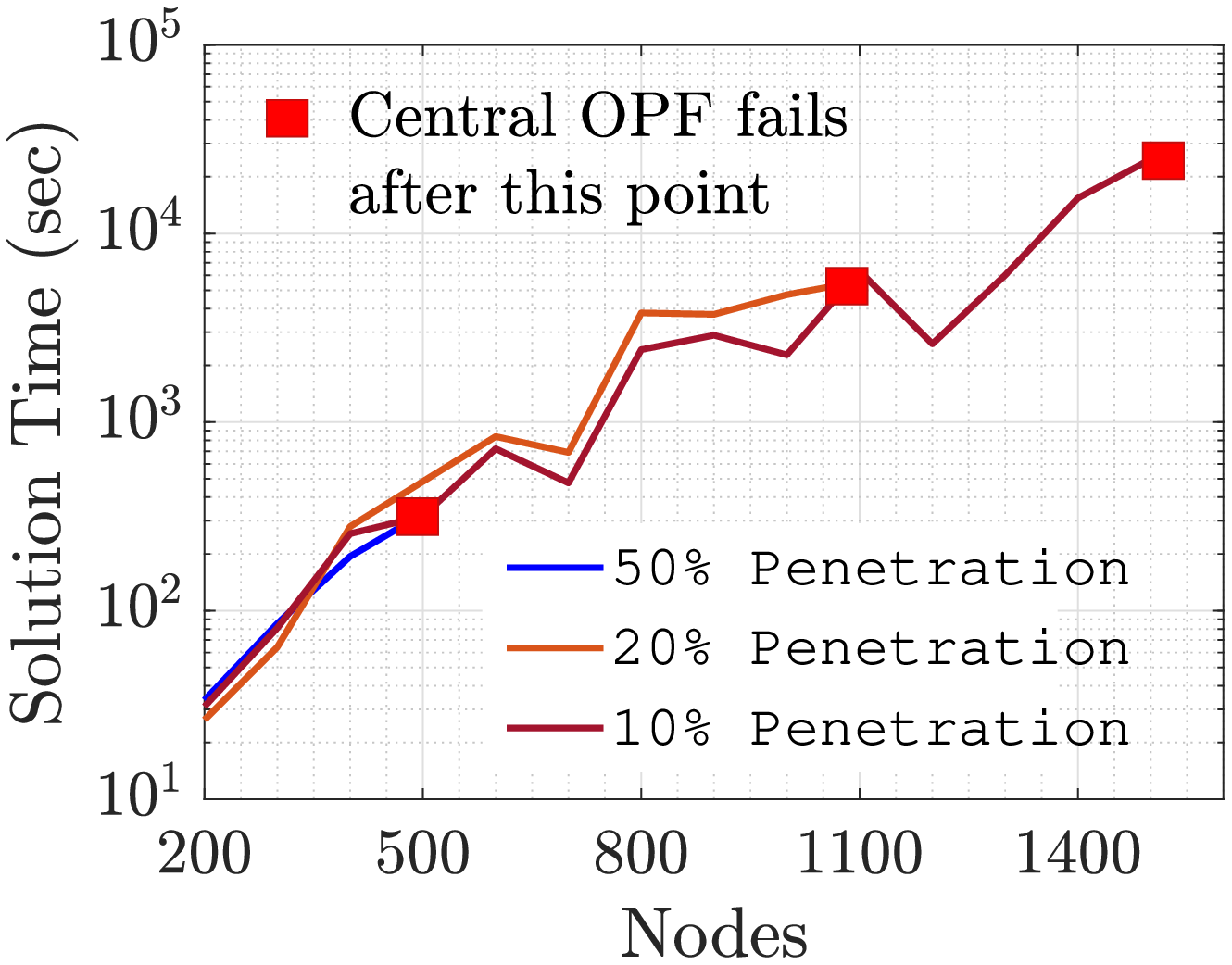}
  \caption{DER maximization objective}
  \label{fail_pm}
\end{subfigure}
\begin{subfigure}{.32\textwidth}
  \centering
  \includegraphics[width=1.00\linewidth]{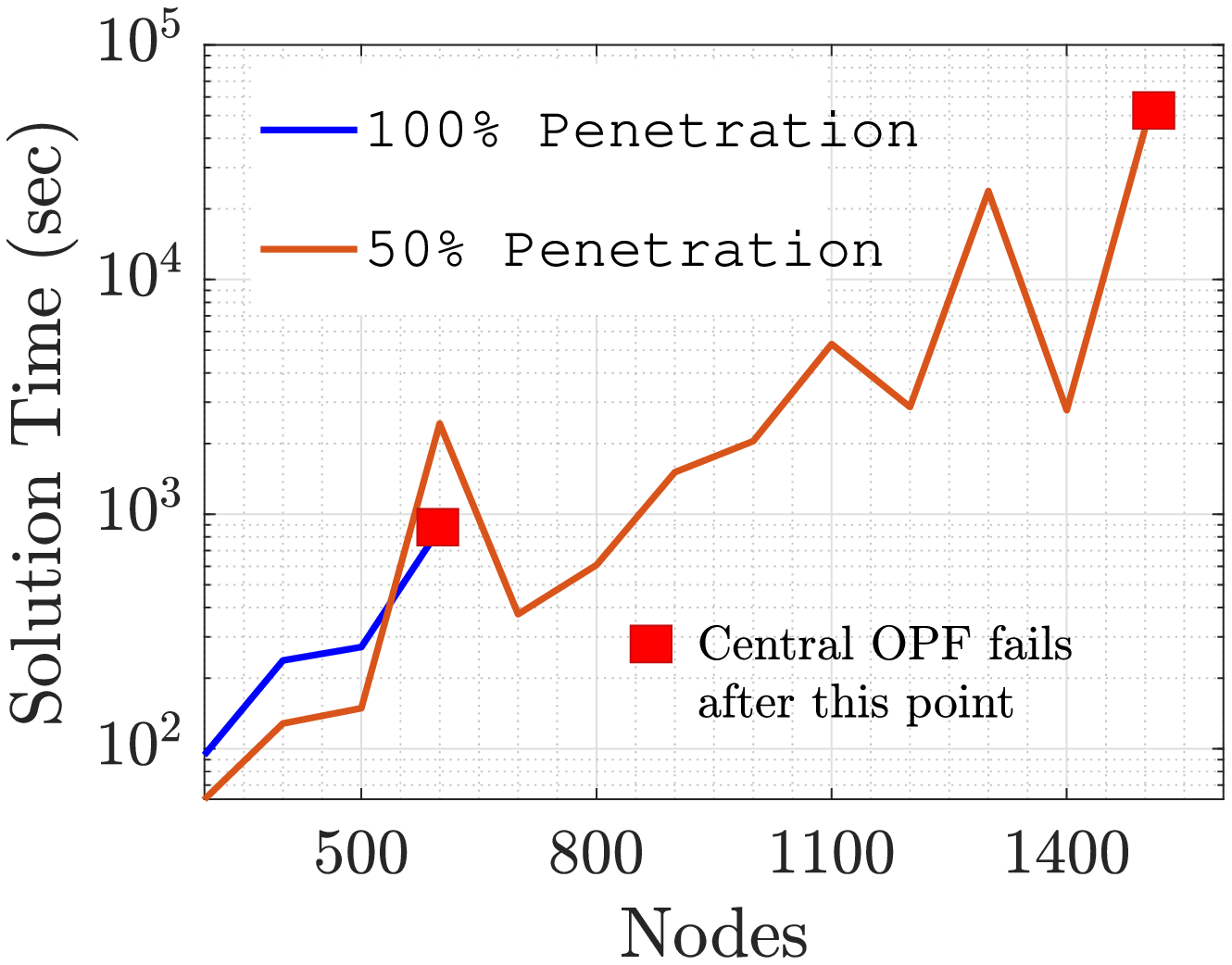}
  \caption{$\Delta$V minimization objective}
  \label{fail_vm}
\end{subfigure}
\caption{Solution Time for Central OPF Problems for Various Objective Functions}
\label{fail_copf}
\end{figure*}

\subsection{Failure of Central Solution}
We have solved the large scale OPFs by decomposing the central problem into several smaller sub-problems, and exchanging the complicating variables. In this section we solve the same central OPF problems for increasing node numbers, and show how poorly the central problem scales with the increasing node numbers. Also, we show when the central OPF fails to solve the OPFs for all the previously simulated cases. From the Fig. \ref{fail_copf}, we can see with increasing DER penetration, it takes higher time to solve the NLP OPFs, and fails earlier. For example, in case of loss minimization, the NLP solver can solve the OPF problems for 800 nodes for 50\% DER penetration case (Fig. \ref{fail_lm}). Similarly, for DER maximization objective, with 20\% DER penetration, the central problem can be solved for no more than 1100 nodes. It is clear that for any OPFs, the NLP problem cannot be solved for a distribution network with more than 2000 nodes. Please note, all of the cases have been solved using knitro with active-set algorithm.

\begin{figure}[t]
    \centering
    \includegraphics[width=0.3\textwidth]{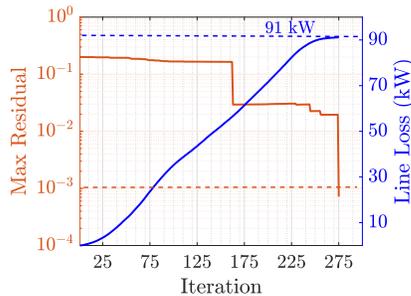}
        % \vspace{-0.2cm}
    \caption{\centering{Result for Nodal Decomposition for IEEE-8500 Node Test System}}
    \label{node_dec}
 \vspace{-0.5cm}
\end{figure}

\subsection{Nodal Decomposition}
To show the efficacy of the algorithm, we further decomposed the problem at individual node level, and solved the OPF problem with loss minimization objective. For this case, we simulated the balanced IEEE-8500 bus test system with 2522 nodes. While the commercial NLP solvers, such as \textit{knitro} fails to optimize this system, the nodal decomposition of the problem speeds up the whole process significantly. It only takes $\sim 5$ seconds to solve the NLP OPF with nodal decomposition of the network. This method decomposes the OPF into 2522 sub-problems, but significantly reduces the variables in each sub-problem -- only 5 variables to solve. It takes 275 macro-iterations to converge, however, since each sub-problem takes microseconds to solve the small NLP problems, the overall solution time is only $\sim 5$ seconds. The convergence is shown in Fig. \ref{node_dec}.

% \subsection{Discussion}

%table:

%%%%  Conclusion: %%%%%%%%%
\section{Conclusion}
In this paper we have solved the large scale OPF problems for power distribution systems by decomposing the networks and distributing the problem into smaller sub-problems. The proposed approach is a generalized decomposition for radial power distribution system that scales very well for all general classes of distributed OPF problems. %The radial operation topology of power distribution systems is leveraged in the decomposition approach, and the assumptions taken to solve each sub-problems stems from the nature of power flows associated with such radial grids.
% From Table \ref{res_tab}, we can see that for different problem objectives, i.e., loss minimization, DER maximization, and $\Delta$V minimization,
The proposed distributed approach converges within fewer iterations and in a short-period of time for large feeders even for the cases where the centralized OPF either takes significant amount of time or simply fails to converge. We have demonstrated the successful application of the proposed approach for a synthetic 10,000 node distribution test system with a total $\sim 50,000$ variables, on a regular CPU, and within reasonable time. 
% In addition to that, due to the nature of decomposition, the DER penetration doesn't have significant influence on the solution time. However, the objective function can have an impact on the solution time. We observed that the DER maximization objective takes longer time than other OPFs. On the other hand, macro-iterations numbers are linearly dependent on the number of decomposed areas. That was consistent for the nodal decomposition case as well (Fig. \ref{node_dec}).
% These have significantly improved the convergence time and in turns, the overall solution time. In this paper, we have shown that the proposed distributed computing method paves the way to solve large scale D-OPF problems with regular CPU.
To the best of our knowledge, this is the first work to demonstrate the application of distributed algorithms to solve OPF problem for the selected very large distribution feeder without requiring HPC machines. Furthermore, the decomposition is amenable for implementation on many-core machines; the fast convergence and fewer communication requirements for the proposed algorithm will lead to significant advancement in solving large-scale OPF problems for active power distribution systems. Authors are working on extending the approach to the three-phase unbalanced D-OPF problems.

\bibliographystyle{ieeetr}
\bibliography{cite}

% \begin{thebibliography}{1}
% \bibitem{Shell}
% M.~Shell, \emph{How to Use the IEEEtran Latex Class}, Latex Archive Contents, \verb+http://www.ieee.org/conferences_events/+ \verb+conferences/publishing/templates.htm+

% \bibitem{IEEEhowto:kopka}
% H.~Kopka and P.~W. Daly, \emph{A Guide to \LaTeX}, 3rd~ed.\hskip 1em plus
%   0.5em minus 0.4em\relax Harlow, England: Addison-Wesley, 1999.

% \end{thebibliography}

% that's all folks
\end{document}